\documentclass[9pt,twocolumn,twoside]{osajnl}

\journal{ol} 

\setboolean{shortarticle}{true}

\title{Widely tunable cavity-enhanced frequency combs}

\author[1]{Myles C. Silfies}
\author[2]{Grzegorz Kowzan}
\author[1]{Yuning Chen}
\author[1]{Neomi Lewis}
\author[3]{Ryan Hou}
\author[4]{Robin Baehre}
\author[4]{Tobias Gross }
\author[1*]{Thomas K. Allison}

\affil[1]{Departments of Physics and Chemistry, Stony Brook University, Stony Brook NY, 11794}
\affil[2]{Institute of Physics, Faculty of Physics, Astronomy and Informatics, Nicolaus Copernicus University, ul. Grudziądzka 5/7, 87-100 Torun, Poland}
\affil[3]{Department of Physics, Columbia University, New York NY, 10027}
\affil[4]{Laseroptik GmbH, Horster Str. 20, 30826 Garbsen, Germany}
\affil[*]{Corresponding author: thomas.allison@stonybrook.edu}

\newcommand{\textss}[1]{\scriptsize \mbox{#1}}

\begin{abstract}
We describe the cavity-enhancement of frequency combs over a wide tuning range of 450-700 nm ($>$ 7900 cm$^{-1}$), covering nearly the entire visible spectrum. Tunable visible frequency combs from a synchronously-pumped optical parametric oscillator are coupled into a 4-mirror, dispersion-managed cavity with a finesse of 600 to 1400. An intracavity absorption path length enhancement greater than 190 is obtained over the entire tuning range, while preserving intracavity spectral bandwidths capable of supporting sub-200 fs pulse durations. These tunable cavity-enhanced frequency combs can find many applications in nonlinear optics and spectroscopy. 
\end{abstract}

\setboolean{displaycopyright}{true}

\begin{document}

\maketitle
\\

\noindent
The cavity enhancement of stabilized ultrafast pulse trains, or frequency combs, first demonstrated around the turn of the century \cite{Jones_PRA2004, Gherman_OptExp2002}, has since been used in many applications. In this technique, successive pulses from the frequency comb are constructively interfered with a circulating intracavity pulse by tuning both the comb's repetition rate ($f_{\textrm{rep}}$) and carrier-envelope offset frequency ($f_0$) such that the frequency comb’s ``teeth” are matched with the enhancement cavity resonance frequencies over a large spectral bandwith. In some applications, this method is used to enhance the intracavity power to drive nonlinear processes at high repetition rate. For example, with kilowatts of circulating average power, one can generate high-order harmonics at high repetition rate \cite{Mills_JPhysB2012}, and this is now being used for precision spectroscopy \cite{Cingoz_Nature2012} and high-repetition rate photoelectron spectroscopy experiments \cite{Corder_StructDyn2018, mills_cavity-enhanced_2019, Saule_NatComm2019}. Other ultrafast nonlinear processes have also used the high intracavity power, such as molecular alignment \cite{Benko_PRL2015} and spontaneous parametric down conversion \cite{Krischek_NatPhot2010}. In another class of applications, it is the enhancement of sensitivity that is sought, such as in cavity-enhanced direct frequency comb spectroscopy, reviewed by Adler et al. in 2010 \cite{Adler_AnnRevChem2010} and continuing to make rapid progress since then \cite{kowzan_broadband_2019, Changala_S_Rovibrational_2019, Alrahman_OEO_Cavityenhanced_2014}. A new application of cavity-enhanced frequency combs developed by our group uses the enhancement of both laser power and sensitivity to obtain a large improvement in the detection-limits of ultrafast nonlinear spectroscopy \cite{Reber_Optica2016, Allison_JPhysB2017}.

In all of this previous work, the cavities have been carefully designed to enhance frequency combs with a certain center wavelength for a specifically targeted intracavity experiment, with limited bandwidth and little or no tuning range. In contrast, frequency comb experiments without cavities are increasingly using combs covering very wide spectral ranges, and the development of widely tunable and broadband frequency combs has recently been the subject of intense research \cite{Schliesser_NatPhot2012,Timmers_Optica2018,Seidel_SciAdv2018,LeinDecker_OptExp2012,Lee_OptLett2013, Ruehl_OptLett2012,Sobon_OptLett2017, Steinle_OptLett2016, Steinle_OptExp2014}. 

The cavity-enhancement of frequency combs over a wide spectral range poses a number of technical challenges. First, one must have a tunable frequency comb with transducers to control both the $f_{\textss{rep}}$ and $f_0$ of the comb, with at least one high-bandwidth transducer to tightly lock the comb to the cavity or vice/versa. This rules out, for example, the recently popular offset-free ``DFG combs'' \cite{Maser_ApplPhysB2017,catanese_mid-infrared_2019}. Second, the combs should have low optical phase noise. For typical repetition rates of $\sim100$ MHz, even modest cavity finesses of $\sim1000$ have small optical linewidths of $\sim100$ kHz. The linewidths of the comb's ``teeth'' should then be substantially narrower than this in order to avoid increased intensity noise on the intracavity light or a reduction in effective input power, and accomplishing this requires special care \cite{Li_RSI2016}. Third, in order to couple a large comb bandwidth, and thus short pulses, into the cavity, very good control of the cavity's intracavity group-delay dispersion (GDD) must be achieved. For example, if a 100 fs input pulse duration is to remain less than 200 fs in a cavity with a finesse of 1000, the net GDD must be less than $\sim100$ fs$^2$. Achieving intracavity GDD this low at one design wavelength is straightforward, but managing dispersion at this level over a wide tuning range, while also satisfying the constraints imposed by high cavity finesse and high average powers requires careful design of the cavity mirror coatings and precise control of the coating process.
\begin{figure}[t]
	\centering
	\includegraphics[width=\linewidth]{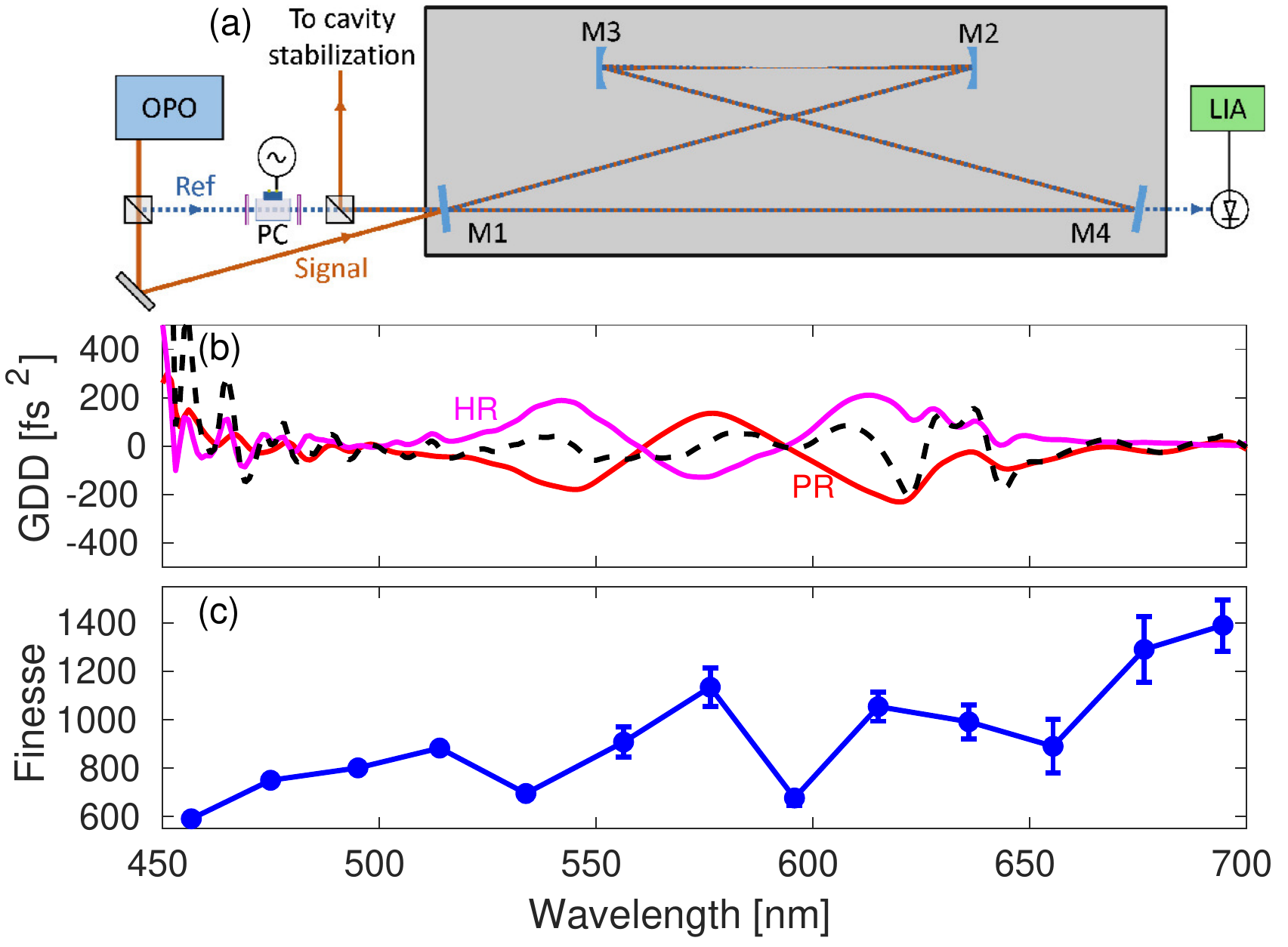}
	\caption{(a) Setup for the CAPS measurement and design of the enhancement cavity. M1 and M4 are partial reflectors (PR) and M2 and M3 are high reflectors (HR). PC: Pockels cell, LIA: lock-in amplifier (b) GDD for the two mirror coatings. Black dashed line is the net cavity GDD assuming the geometry in (a). (c) Measured cavity finesse.}
	\label{fig:F_GDD}
\end{figure}

In this letter, we address these challenges, focusing on the visible spectral range, using a recently developed widely tunable frequency comb source \cite{Chen_ApplPhysB2019} and a novel cavity coating design. The design of our femtosecond enhancement cavity (fsEC) is shown in Fig. \ref{fig:F_GDD}(a). The fsEC is a 4-mirror bow-tie configuration with 2 plane partial reflectors, M1 and M4, of nominally 0.3\% transmission and 2 high reflectors (transmission < 0.1\%), M2 and M3, with 50 cm radius of curvature. The mirrors are designed for operation from 450 to 700 nm. In order to minimize net cavity GDD, the two mirror coatings have opposite chirp i.e., the high reflector is constructed such that longer wavelengths reflect from deeper in the coating whereas in the partial reflector this relation is reversed \cite{kaertner_broadband_2011}. The GDD, measured by white light interferometry, is shown in Fig. \ref{fig:F_GDD}(b) for both mirrors as well as the expected net cavity GDD assuming two of each mirror (black, dashed). Both the net GDD and reflectivity are varying and structured which is expected for such a large operating range, and this results in an overall cavity performance that varies with wavelength. The mirrors were coated using ion beam sputtering with alternating layers of Ta$_2$O$_5$ and SiO$_2$. The total coating structures have thickness of 6.45 $\mu$m and 4.86 $\mu$m for the HR and PR respectively and the coatings were annealed first at $200^{\circ}$C  for 5 hours and then at $300^{\circ}$C for an additional 4 hours to reduce loss. The spot size at the smaller waist is calculated using the ABCD matrix formalism to have a 1/$e^2$ radius of 65 \textmu m at 530 nm. The spot size scales with wavelength only weakly as $\sqrt{\lambda}$ \cite{Siegman_book1986}. The entire cavity is contained in a vacuum chamber which is held at a pressure below 100 mTorr to minimize GDD from air and avoid air currents.

To characterize the fsEC performance, we first measure the cavity finesse, $\mathcal{F}$, across the design wavelength range. To measure the finesse independent of the complications of the comb cavity coupling described below, we use the cavity attenuated phase shift (CAPS) method \cite{herbelin_sensitive_1980,engeln_phase_1996}. In this method, the input laser to the cavity is amplitude modulated at an angular frequency $\Omega$ and the relative phase shift of the cavity transmitted light is measured. The phase shift, $\phi$, is due to the cavity acting as a low pass filter and is related to the cavity storage time, $\tau$, by:
\begin{equation}
\label{eqn_phi}
\mathrm{tan}(\phi) = -\Omega\tau
\end{equation}
Which, in turn, is related to finesse via:
\begin{equation}
\label{eqn_finesse}
\mathcal{F} = 2\pi \tau f_{\textrm{rep}}
\end{equation}

To measure finesse via the CAPS method while remaining frequency-locked to the cavity, we send two identical beams into the cavity in counterpropogating directions as shown in Fig. \ref{fig:F_GDD}(a) and described in \cite{Reber_Optica2016}. The forward beam, labeled signal, is used for locking while the second, labeled reference, is used for the measurement. For our CAPS implementation, the reference beam is sent through a Pockels cell between two polarizers. The Pockels cell is driven with sine wave with a peak value $\approx$ 3\% of the half-wave voltage. The first polarizer is detuned from the second, which optically biases the cell to get a cleaner sinusoidal output at the drive frequency. The reference beam at the cavity output is sent to a photodiode and lock-in amplifier measuring the phase shift relative to a leakage reference beam which bypasses the cavity. 

The measured cavity finesse across the design range is shown in Fig. \ref{fig:F_GDD}(c). The mean value is 927. The data shown is an average of 2 CAPS phase measurements taken at 70 and 90 kHz modulation frequency. At each frequency, the lock-in phase is recorded for 10 seconds and then the average and standard deviation is calculated which is used to find the error in each measurement. The error is higher on the long wavelength side due to both lower output intensity and higher intensity noise, which will be discussed below. As expected, the cavity finesse varies by more than a factor of 2 across the design range, with a general trend of increasing finesse at longer wavelengths. This is most likely due to absorption of the bluer side of the spectrum in the mirror coatings. The original design specification was for a finesse of $\sim$ 1000 which is satisfied for most of the range. The cavity loss is dominated by the two PR mirrors and the cavity is nearly impedance matched \cite{nagourney_quantum_2014}. In this case the absorption sensitivity enhancement factor remains approximately $\mathcal{F}/\pi$ and is greater than 190 across the tuning range \cite{gagliardi_cavity-enhanced_2014}.

 \begin{figure}[t]
 	\centering
 	\includegraphics[width=\linewidth]{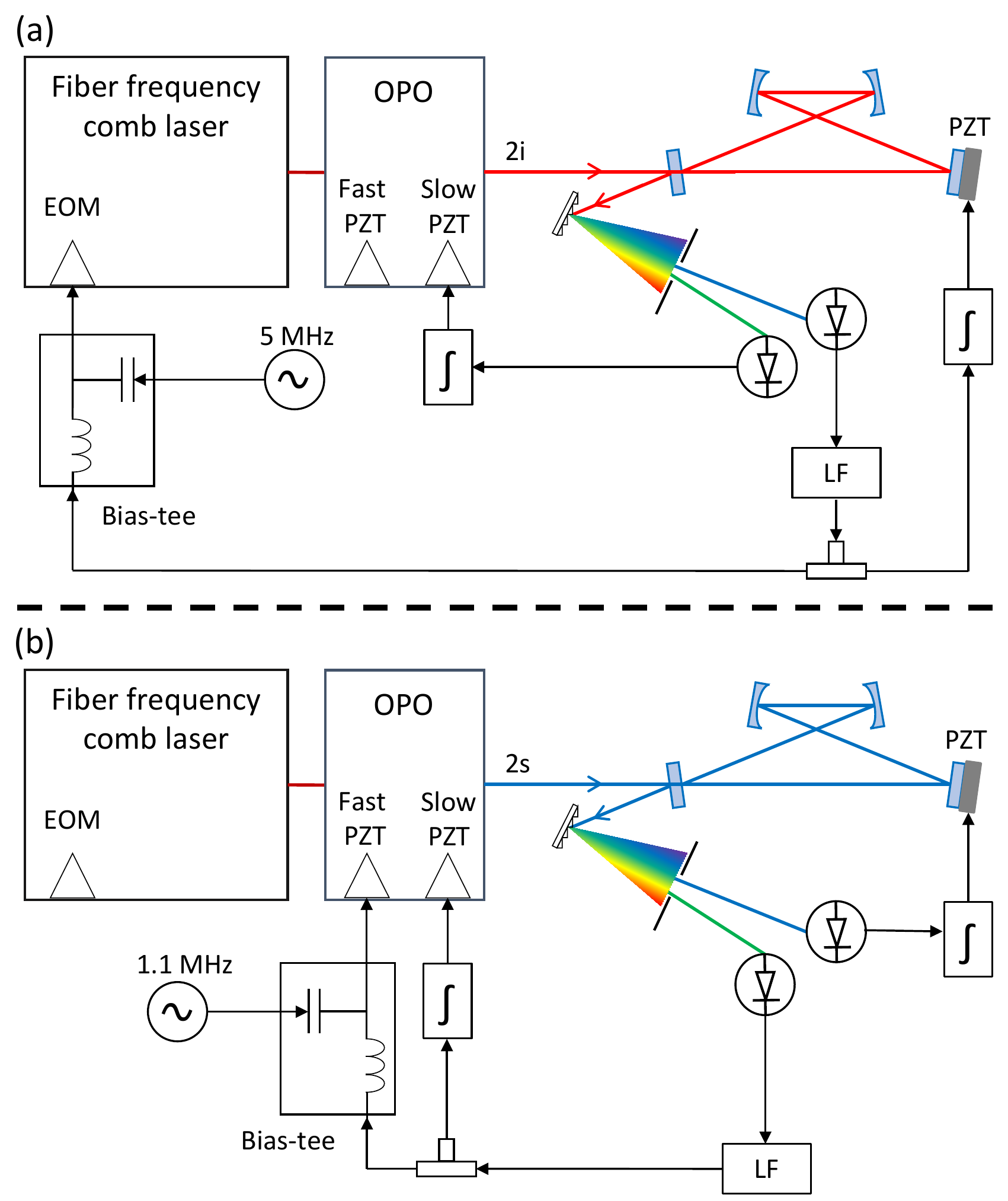}
 	\caption{Schematics for locking input comb to the enhancement cavity for (a) the doubled idler and (b) the doubled signal. PZT: piezoelectric transducer, EOM: electro-optic modulator, LF: loop filter}
 	\label{fig:cav_lock}
 \end{figure}

The CAPS measurements are relatively simple and independent of the details of the comb/cavity coupling. For measuring the performance of the setup for cavity-enhancing frequency combs with maximum bandwidth, more discussion of the frequency comb and stabilization schemes is required. For all these measurements, we use a two-point Pound-Drever-Hall (PDH) locking scheme \cite{Drever_ApplPhysB1983}, with a fast servo loop tightly locking one portion of the frequency comb to the cavity, and a second slower loop bringing another part of the comb onto resonance with a linearly independent actuator. This basic scheme has been used previously in many contexts \cite{Jones_PRA2004, Foltynowicz_APB2013, Reber_Optica2016, Corder_SPIE2018}. However, accomplishing comb/cavity coupling over such a large and unprecedented tuning range presents technical challenges, which we discuss in detail below.  

The input combs are derived from a home-built synchronously pumped optical parametric oscillator (OPO) operating at 100 MHz repetition rate, previously described in \cite{Chen_ApplPhysB2019}. This OPO is pumped by the second harmonic of a home-built high-power frequency comb laser consisting of a dispersive-wave shifted Er:fiber comb \cite{Maser_ApplPhysB2017} with high-bandwidth transducers, amplified in a Yb-doped photonic crystal fiber amplifier \cite{Li_RSI2016}. To cover the entire tuning range, we use all three of the frequency doubled signal ($2s$, 450-515 nm), residual pump (535 nm), and doubled idler ($2i$, 555-700 nm) from the OPO. The phase transfer properties of the OPO, studied in detail in \cite{Chen_ApplPhysB2019}, necessitate different comb/cavity stabilization schemes for each of these combs. The schemes for $2i$ and $2s$ combs are shown in Fig. \ref{fig:cav_lock}.

For coupling the $2i$ comb to the cavity, since phase modulation on the pump is transferred directly to the $2i$ comb with no bandwidth penalty imposed by the OPO cavity \cite{Chen_ApplPhysB2019}, we use an electro-optic modulator (EOM) in the mode-locked fiber laser to apply both 5 MHz PDH sidebands and for feedback in the fast PDH loop, as shown in Fig. \ref{fig:cav_lock}(a). This acts on the pump and $2i$ combs with a fixed point \cite{Newbury_JOSAB2007} near DC, mainly acting on the $f_{\textss{rep}}$ degree of freedom of the combs. The cavity's free spectral range is kept within the EOM's $f_{\textss{frep}}$ tuning range with an additional slow servo acting on the enhancement cavity length. For controlling the second comb degree of freedom in the slow PDH loop, we actuate on the OPO cavity length, which acts only on the comb's $f_0$ degree of freedom. When coupling the pump light to the enhancement cavity, the fast loop is identical to the $2i$ case shown in Fig. \ref{fig:cav_lock}(a) and the slow loop actuates on a the temperature of the EOM in the mode-locked fiber laser to control $f_0$. 

For coupling the $2s$ comb to the cavity, the high-speed actuators of the pump laser are of no use since phase modulation on the pump comb is not transferred to the $2s$ comb, as discussed in detail in \cite{Chen_ApplPhysB2019}. Even though $f_{rep}$ of the $2s$ comb must track that of the pump comb, these changes are offset by changes in the $2s$ comb's $f_0$ such that the fixed point of the $2s$ comb with respect to pump phase modulation is near the $2s$ comb's optical carrier frequency. Thus, comb/cavity coupling is accomplished with the pump comb free-running, and feedback signals are instead applied to the OPO cavity and fsEC, as shown in Fig. \ref{fig:cav_lock})(b) The fast servo PDH loop drives a copper-bullet-style piezoelectric transducer (PZT) \cite{Briles_OptExp2010} to actuate on OPO cavity length and the $f_0$ of the $2s$ comb. PDH sidebands at 1.1 MHz are also generated by this PZT at a mechanical resonance. The slow loop controls the fsEC cavity length with a long travel PZT. When locking either the $2i$ or $2s$ combs, care must be taken when selecting the locking points in the spectrum since small changes in OPO cavity length can change the output spectrum dramatically which can result in complicated, multi-peaked intracavity spectra or significantly decreased power. Monitoring of the input and intracavity spectra simultaneously while locking helps avoid this problem. 

\begin{figure}
	\centering
	\includegraphics[width=\linewidth]{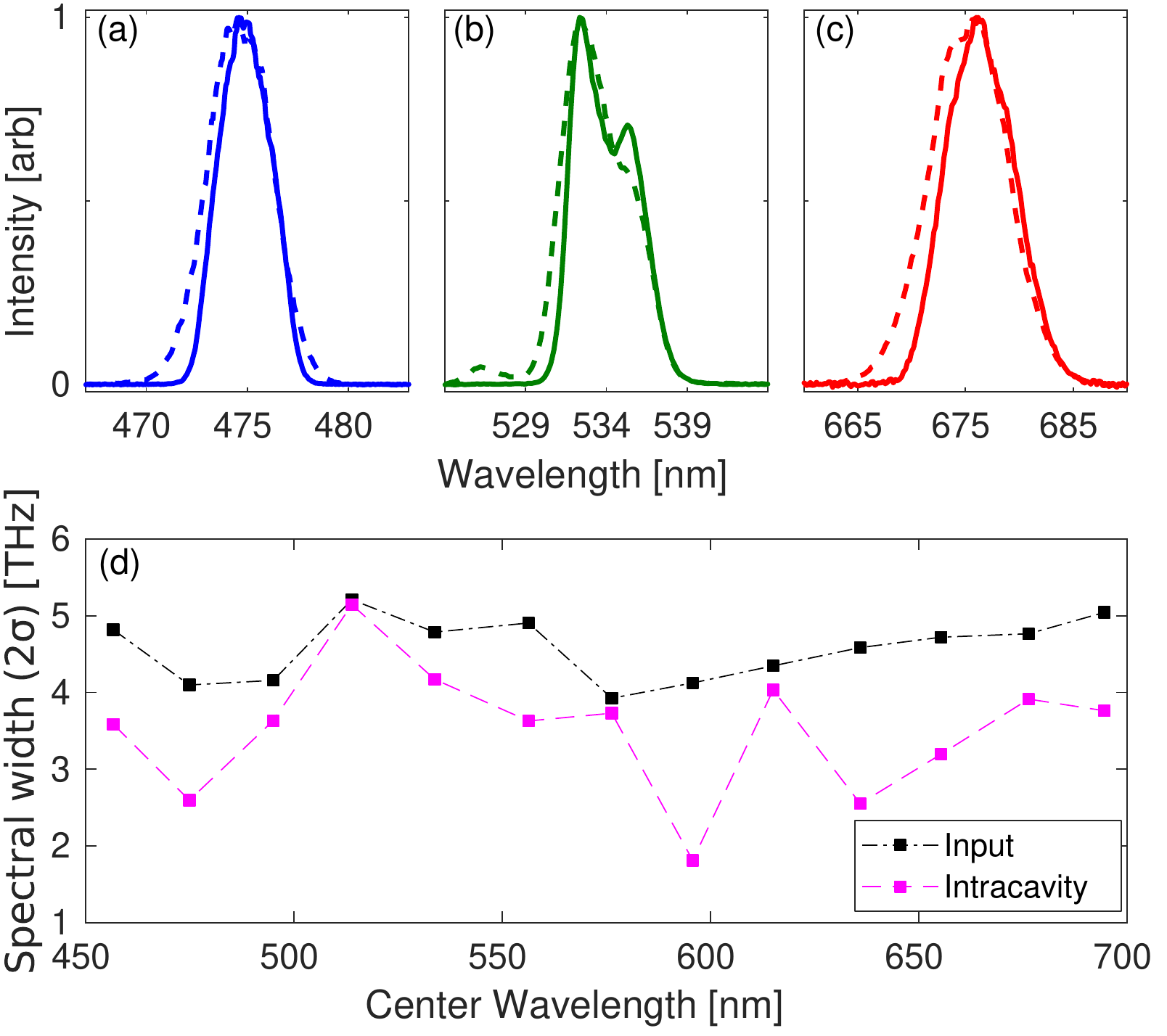}
	\caption{(a-c) Intracavity (solid) and input (dashed) spectra for $2s$ (a), pump (b), and $2i$ (c) combs. (d) Optimized intracavity spectral widths compared with the incident width.}
	\label{fig:T_spec}
\end{figure}

\begin{figure}[b!]
	\centering
	\includegraphics[width=\linewidth]{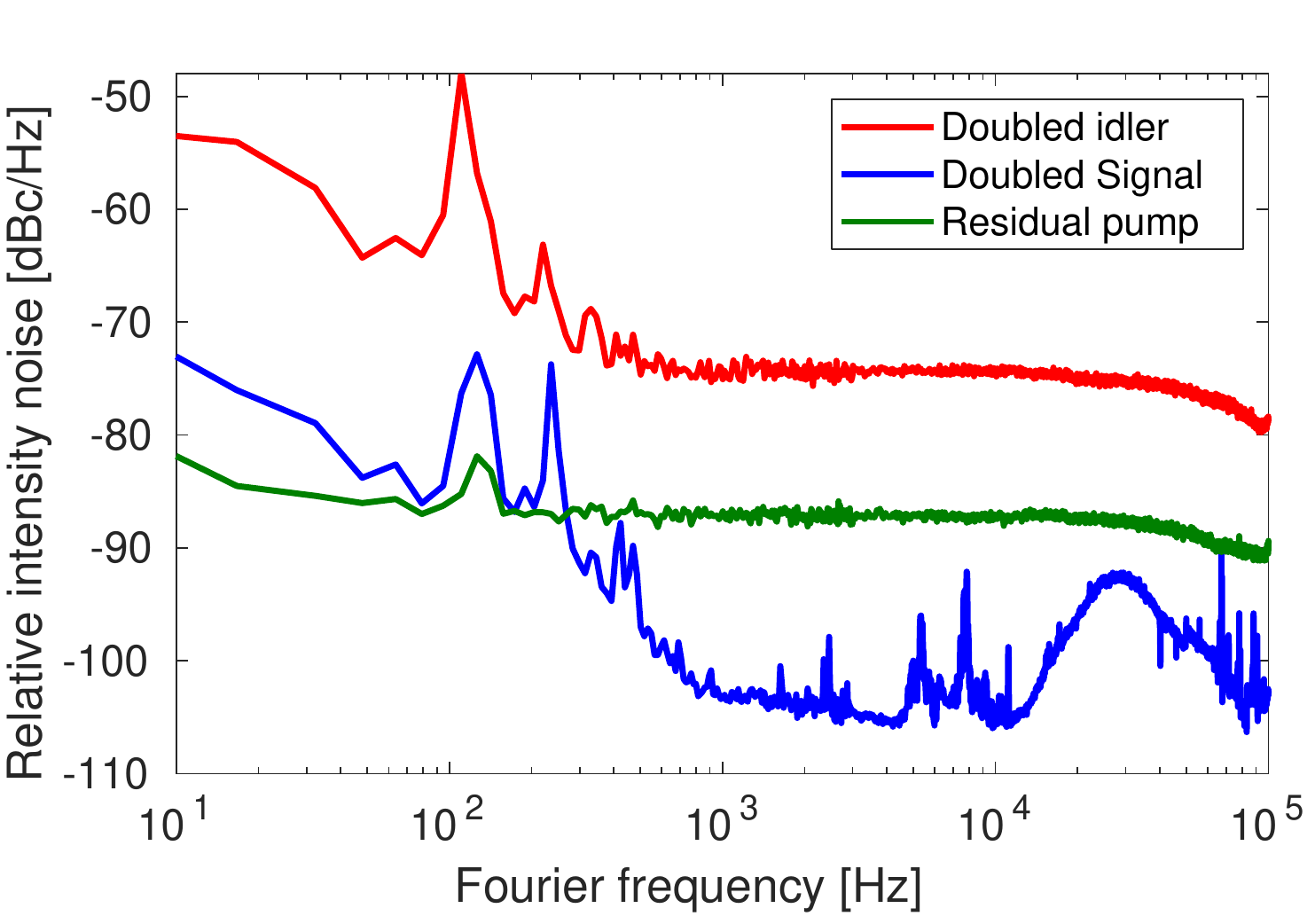}
	\caption{Relative intensity noise of the intracavity light.}
	\label{fig:RIN}
\end{figure}

The fundamental limit to the attainable intracavity pulse duration is given by the attainable intracavity spectral width, which is in turn related to the input spectrum and the intracavity dispersion. We have measured both the incident OPO input spectrum and intracavity spectra across the tuning range. Example intracavity spectra of the $2s$, $2i$, and pump combs are shown in Figs. \ref{fig:T_spec}(a)-(c), along with the corresponding input spectra. Figure \ref{fig:T_spec}(d) shows the input and intracavity spectral widths (2$\sigma$) measured at 13 wavelengths across the tuning range. The attainable intracavity bandwidth is less than the input OPO bandwidth due to residual mirror dispersion. Due to GDD, the intracavity spectra are more square-shaped than the input so standard Gaussian transform limit relations do not apply. Still, even the smallest bandwidths measured support sub-200 fs FWHM pulses across the tuning range when Fourier transformed.

In order to use the cavity-enhanced combs for our target application of ultrasensitive ultrafast optical spectroscopy \cite{Reber_Optica2016, Allison_JPhysB2017}, low noise on the intracavity light is required. We have previously measured the amplitude noise of the input combs and found it to be sufficient \cite{Chen_ApplPhysB2019}. However, since an optical cavity is a phase discriminator \cite{nagourney_quantum_2014,zhu_JOSAB_Stabilization_1993}, residual phase noise present on the input comb is converted to amplitude noise on the intracavity light. The data is shown in Fig. \ref{fig:RIN} for the $2s$, $2i$, and pump combs. The OPO idler inherits the phase noise from the pump and signal such that the $2i$ comb has the highest phase noise \cite{Chen_ApplPhysB2019} of the three combs and thus the largest high-frequency RIN. Conversely, the $2s$ comb has the lowest optical phase noise and thus the lowest high-frequency RIN, despite far inferior servo bandwidth for the PZT locking scheme of Fig. \ref{fig:cav_lock}(b).  

In conclusion, we have for the first time demonstrated the cavity-enhancement of frequency combs with a widely tunable platform. The wide tuning range of $> 7900$ cm$^{-1}$ covers nearly the entire visible spectrum. Despite the technical challenges wavelength tuning imposes on the frequency comb generation, cavity mirrors, and comb/cavity coupling, we have demonstrated performance that is comparable to that used in previous experiments using cavity-enhanced combs \cite{Reber_Optica2016, Corder_StructDyn2018, Lilienfein_NatPhot2019}. For example, comparing to the previous one-wavelength demonstration of cavity-enhanced transient absorption spectroscopy \cite{Reber_Optica2016}, we report here the achievement of comparable or better intracavity RIN across the tuning range, with higher cavity finesse, while maintaining similar intracavity optical bandwidth. We expect further refinements in cavity-mirror coating technology and frequency combs to further improve the attainable performance.
\\
\\
\noindent\textbf{Funding.} National Science Foundation (NSF) (1708743). Air Force Office of Scientific Research (AFOSR) (FA9550-16-1-0164).
\\
\\
\noindent\textbf{Acknowledgments.} M.C. Silfies acknowledges support from the GAANN program of the U.S. Dept. of Education. G. Kowzan acknowledges support from the National Science Centre, Poland scholarship 2017/24/T/ST2/00242. The authors thank S.A. Diddams, H. Timmers, A. Kowligy, A. Lind, F.C. Cruz, N. Nader, and G. Ycas for assistance with Er comb development and providing manuscript feedback.
\\
\\
\noindent\textbf{Disclosures.} The authors declare no conflicts of interest.

\end{document}